\documentclass[prd,twocolumn,showpacs,preprintnumbers,amsmath,amssymb]{revtex4}
\newcommand{\re}{\mathrm{Re}}
\newcommand{\im}{\mathrm{Im}}

\newcommand{\half}{{{}^1_{\overline 2}}}

\begin{document}

\title{Polynomial Interpretation of Multipole Vectors}

\author{Gabriel Katz}
\affiliation{Bennington College, Bennington VT 05201 USA}%
\email{gkatz@bennington.edu}

\author{Jeff Weeks}
\affiliation{15 Farmer Street, Canton NY 13617 USA}%
\email{www.geometrygames.org/contact.html}

\date{\today}

\begin{abstract}
\noindent Copi, Huterer, Starkman and Schwarz introduced multipole
vectors in a tensor context and used them to demonstrate that the
first-year WMAP quadrupole and octopole planes align at roughly
the 99.9\% confidence level.  In the present article the language
of polynomials provides a new and independent derivation of the
multipole vector concept.  B\'ezout's Theorem supports an
elementary proof that the multipole vectors exist and are unique
(up to rescaling). The constructive nature of the proof leads to a
fast, practical algorithm for computing multipole vectors.  We
illustrate the algorithm by finding exact solutions for some
simple toy examples, and numerical solutions for the first-year
WMAP quadrupole and octopole.  We then apply our algorithm to
Monte Carlo skies to independently re-confirm the estimate that
the WMAP quadrupole and octopole planes align at the 99.9\% level.
\end{abstract}


\maketitle

\section{Introduction}

The first-year WMAP data \cite{WMAP-Bennett} reveal a somewhat
planar octopole which approximately aligns with the quadrupole
\cite{Tegmark}.  More recent studies confirm these conclusions at
roughly the 99.9\% level \footnote{The preprint \cite{Schwarz}
states a 99.97\% confidence level, but contains an error in the
statistical analysis. Correcting the error reduces the confidence
level to 99.9\%, which is of course still excellent.} while
revealing mysterious alignments with the ecliptic plane
\cite{Schwarz}, suggesting either a hitherto unknown solar system
effect on the microwave background or an error in the collection
and/or processing of the WMAP data. Other researchers find the
$\ell = 4$ multipole is generic, the $\ell = 5$ multipole is
unusually {\it non}-planar at the 99.8\% level, and the $\ell = 6$
multipole is unusually planar at the 98.6\% level \cite{Eriksen}.
No explanation is yet known for these strange results.

{\it Multipole vectors} provide a convenient means to quantify the
planarity of a given multipole, as well as to compare the
alignment of two different multipoles \cite{Copi}.  The present
authors, coming from a background in pure mathematics, were unable
to decipher the formalism and terminology of Ref.~\cite{Copi} and
chose instead to re-create the multipole vector concept from
scratch.  The real-valued spherical harmonics of order $\ell$ are
precisely the homogeneous harmonic polynomials of degree $\ell$ in
the variables $x$, $y$ and $z$ (for example $Y_{2}^{0}$ is the
polynomial $x^2 + y^2 - 2 z^2$, up to normalization), so the
present authors sought to understand the multipole vectors of
Copi, Huterer and Starkman (CHS) from a polynomial point of view.

Translated to the language of polynomials, CHS's motivating goal
(see Eqn.~(10) of \cite{Copi}) was to factor every homogeneous
harmonic polynomial $P$ of degree $\ell$ into a product of linear
factors
\begin{eqnarray}
  P(x,y,z) = \lambda &\cdot& (a_1 x + b_1 y + c_1 z) \nonumber\\
                     &\cdot& (a_2 x + b_2 y + c_2 z) \nonumber\\
                     &\phantom{\cdot}&\cdots         \nonumber\\
                     &\cdot&(a_\ell x + b_\ell y + c_\ell z).
\end{eqnarray}
Such a factorization is of course impossible in general, as CHS
implicitly acknowledge by their introduction of suitable error
terms.  In the language of polynomials the correct statement of
the theorem is

\vskip0.25cm

\noindent {\it {\bf Theorem 1.}  Every homogeneous polynomial $P$
of degree $\ell$ in $x$, $y$ and $z$ may be written as}
\begin{eqnarray}
\label{MainTheoremEquation}
  P(x,y,z) &=& \lambda \cdot (a_1 x + b_1 y + c_1 z)                 \nonumber\\
           & & \phantom{\lambda}\cdot(a_2 x + b_2 y + c_2 z)         \nonumber\\
           & & \phantom{\lambda\cdot}\quad\cdots                     \nonumber\\
           & & \phantom{\lambda}\cdot(a_\ell x + b_\ell y + c_\ell z)\nonumber\\
           &+& (x^2 + y^2 + z^2)\cdot R,
\end{eqnarray}
{\it where the remainder term {\rm R} is a homogeneous polynomial
of degree $\ell - 2$.  The decomposition is unique up to
reordering and rescaling the linear factors.}

\vskip0.25cm

\noindent Notes:  {\it (a)}  Theorem 1 lives entirely in the realm
of real polynomials:  the coefficients of $P$, $R$ and all the
linear factors $a_i x + b_i y + c_i z$ are assumed to be real.
{\it (b)}  Theorem 1 does not require the polynomial $P$ to be
harmonic.

\vskip0.25cm

In cosmological applications we are interested only in the value
of the polynomial on the unit sphere $S^2$;  we ignore its value
on the rest of Euclidean 3-space.  On the unit sphere the factor
$x^2 + y^2 + z^2$ is identically 1, so in this case Theorem~1 says
that any homogeneous polynomial $P$ may be written as a product of
linear factors $\lambda (a_1 x + b_1 y + c_1 z)\cdots(a_\ell x +
b_\ell y + c_\ell z)$ plus a remainder term $R$ of lower degree.
Applying this reasoning recursively gives the easy

\vskip0.5cm

\noindent {\it {\bf Corollary 2.}  When restricted to the unit
sphere, every polynomial $P$ of degree $\ell$ in $x$, $y$ and $z$
may be written as P(x,y,z) =}
\begin{eqnarray}
\label{Corollary2}
  &\phantom{+}& \lambda_{\ell} \cdot (a_{\ell,1} x + b_{\ell,1} y + c_{\ell,1} z)
                    \cdot(a_{\ell,2} x + b_{\ell,2} y + c_{\ell,2} z)\nonumber\\
  &\phantom{+}&\phantom{\lambda_{\ell} \cdot}\qquad\cdots (a_{\ell,\ell} x + b_{\ell,\ell} y + c_{\ell,\ell} z)\nonumber\\
           &+& \dots\nonumber\\
           &+& \lambda_{2} \cdot (a_{2,1} x + b_{2,1} y + c_{2,1} z)\cdot
                     (a_{2,2} x + b_{2,2} y + c_{2,2} z)\nonumber\\
           &+& \lambda_{1} \cdot (a_{1,1} x + b_{1,1} y + c_{1,1} z)\nonumber\\
           &+& \lambda_{0}.
\end{eqnarray}
{\it The decomposition is unique up to reordering and rescaling
the linear factors within each term.}

\vskip0.25cm

\noindent Note:  Corollary 2 does not require the polynomial $P$
to be either homogeneous or harmonic.

\vskip0.25cm

\noindent {\it Proof of Corollary 2.}  Write $P$ as a sum of
homogeneous terms $P = P_{\ell} + P_{\ell - 1} + \dots + P_1 +
P_0$.  First apply Theorem~1 to the highest order term $P_{\ell}$,
yielding a factorization $\lambda_{\ell} \cdot (a_{\ell,1} x +
b_{\ell,1} y + c_{\ell,1} z)\cdots (a_{\ell,\ell} x +
b_{\ell,\ell} y + c_{\ell,\ell} z)$ along with a remainder term
$R_{\ell - 2}$ of homogeneous degree $\ell - 2$.  (The factor $x^2
+ y^2 + z^2$ may be ignored on the unit sphere.)  Fold $R_{\ell -
2}$ in with $P_{\ell - 2}$, and proceed recursively, applying
Theorem~1 to $P_{\ell - 1}$, then $P_{\ell - 2}$, and so on.

To prove uniqueness, consider the even and odd parts of $P$
separately.  That is, write $P = P_{\mathrm{even}} +
P_{\mathrm{odd}}$, where $P_{\mathrm{even}}$ contains all the
even-powered terms and $P_{\mathrm{odd}}$ contains all the
odd-powered terms. Say we have two potentially different
decompositions for the even part
\begin{eqnarray}
  P_{\mathrm{even}}
  &=& \Pi_{\ell}  + \Pi_{\ell - 2}  + \Pi_{\ell - 4}  + \dots + \Pi_0\nonumber\\
  &=& \Pi_{\ell}' + \Pi_{\ell - 2}' + \Pi_{\ell - 4}' + \dots + \Pi_0'.
\end{eqnarray}
where each $\Pi_i$ is the $i^{th}$ term in a decomposition
(\ref{Corollary2}), and where the leading index will be $\ell$ or
$\ell - 1$ according to whether $\ell$ is even or odd. To make
these decompositions homogeneous, multiply through by appropriate
powers of $Q = x^2 + y^2 + z^2$,
\begin{eqnarray}
  P_{\mathrm{even}}
  &=& \Pi_{\ell}  + Q \Pi_{\ell - 2}  + Q^2 \Pi_{\ell - 4}
       + \dots + Q^{\ell/2} \Pi_0\nonumber\\
  &=& \Pi_{\ell}' + Q \Pi_{\ell - 2}' + Q^2 \Pi_{\ell - 4}'
       + \dots + Q^{\ell/2} \Pi_0'.
\end{eqnarray}
This does not affect the value of $P$ on the unit sphere, because
$Q = 1$ there.  The uniqueness part of Theorem~1 implies that the
leading order terms $\Pi_{\ell}$ and $\Pi_{\ell}'$ must be equal.
So subtract off those leading terms, divide through by $Q$, and
apply Theorem~1 again to conclude $\Pi_{\ell - 2} = \Pi_{\ell -
2}'$.  Continue recursively to finally reach $\Pi_{0} = \Pi_{0}'$.
The same argument then proves that the odd part of $P$ has a
unique decomposition as well.  Q.E.D.

\section{Proof of the Main Theorem}
\label{SectionProof}

Even though the statement of Theorem 1 lives wholly in the world
of real polynomials, its proof will dive deeply into the world of
complex polynomials.  So let the variables $x$, $y$ and $z$ range
over the complex numbers, while insisting that the coefficients of
the polynomial $P$ remain real. Because $P$ has homogeneous degree
$\ell$, whenever one point $(x_0, y_0, z_0)$ satisfies $P(x,y,z) =
0$, every nonzero constant multiple $(\alpha x_0, \alpha y_0,
\alpha z_0)$ satisfies it as well.  Thus the equation $P = 0$ is
well defined on each equivalence class of points $\{ \alpha (x_0,
y_0, z_0)\;|\;\alpha \in \mathbb{C} - \{0\} \}$.  In other words,
the complex curve $P = 0$ is well defined on the complex
projective plane $\mathbb{C}P^2$, which is the quotient of
$\mathbb{C}^3 - \{(0,0,0)\}$ under the equivalence relation $(x_0,
y_0, z_0)$ $\sim$ $\alpha (x_0, y_0, z_0)$.  This leads us into
the realm of algebraic geometry and puts its powerful tools at our
disposal.

The most useful tool for our purposes is

\vskip0.25cm

\noindent {\it {\bf B\'ezout's Theorem.}  If $P$ and $Q$ are
homogeneous polynomials of degree $m$ and $n$, respectively, then
the curves $P = 0$ and $Q = 0$ intersect in $\mathbb{C}P^2$
\begin{itemize}
  \item in exactly $m \cdot n$ points, counted with multiplicity,
          if $P$ and $Q$ share no common factor, or
  \item in infinitely many points,
          if $P$ and $Q$ do share a common factor.
\end{itemize}
} %
\noindent For an elementary exposition of B\'ezout's Theorem, see
\cite{Reid}.

\vskip0.25cm

\noindent In the present case, the only way the polynomial $P$ may
share a factor with the irreducible polynomial $Q(x,y,z) \equiv
x^2 + y^2 + z^2$ is for $P$ to contain $Q$ as a factor, in which
case Theorem~1 is trivially satisfied (take $\lambda = 0$).  So
henceforth assume $P$ does not contain $Q$ as a factor. B\'ezout's
Theorem now tells us that the degree $\ell$ complex curve
$P(x,y,z) = 0$ intersects the quadratic curve $Q(x,y,z) = 0$ in
exactly $2\ell$ points, counted with multiplicities. None of the
intersection points may be purely real, because real values cannot
possibly satisfy $x^2 + y^2 + z^2 = 0$ -- recall that the
definition of $\mathbb{C}P^2$ explicitly excludes $(0,0,0)$.
Furthermore, because $P$ and $Q$ both have real coefficients,
whenever $(x_0, y_0, z_0)$ lies in the intersection $P = Q = 0$,
its complex conjugate $(\overline{x_0}, \overline{y_0},
\overline{z_0})$ must lie there too.  So the $2\ell$ points of
intersection consist of $\ell$ pairs of non-real complex
conjugates $\{p_1, \overline{p_1}, \dots, p_\ell,
\overline{p_\ell}\}$.

We claim that each pair $\{p_i, \overline{p_i}\}$ determines a
unique line $a_i x + b_i y + c_i z = 0$ with real coefficients.
The proof is easy.  The conjugate pair $\{p_i, \overline{p_i}\}$
lies on the line $a_i x + b_i y + c_i z = 0$ if and only if the
real and imaginary parts
satisfy the following two totally real equations
\begin{eqnarray}
\label{LineEquations}
  a_i \; \re\,p_{i,x} + b_i \; \re\,p_{i,y} + c_i \; \re\,p_{i,z} = 0\phantom{.}\nonumber\\
  a_i \; \im\,p_{i,x} + b_i \; \im\,p_{i,y} + c_i \; \im\,p_{i,z} = 0.
\end{eqnarray}
Geometrically those two equations represent planes in $R^3$. If
the coefficient vectors $(\re\,p_{i,x}, \re\,p_{i,y},
\re\,p_{i,z})$ and $(\im\,p_{i,x}, \im\,p_{i,y}, \im\,p_{i,z})$
are non-collinear, then the two planes are distinct and their
intersection, which defines the solution set for $(a_i, b_i,
c_i)$, is a line through the origin in $R^3$.  In other words, the
line $a_i x + b_i y + c_i z = 0$ is unique.  Normalize the
coefficients to unit length, i.e. $a_i^2 + b_i^2 + c_i^2 = 1$, and
the only remaining ambiguity is an overall factor of $\pm 1$.

But what if the coefficient vectors $\re\,p_i$ = $(\re\,p_{i,x}$,
$\re\,p_{i,y}$, $\re\,p_{i,z})$ and $\im\,p_i$ = $(\im\,p_{i,x}$,
$\im\,p_{i,y}$, $\im\,p_{i,z})$ had been collinear?  In this case
the line $a_i x + b_i y + c_i z = 0$ would be ill-defined.
Fortunately this case does not arise. For if $\im\,p_i$ were
proportional to $\re\,p_i$, say $\im\,p_i = \beta\,\re\,p_i$, then
the point $p_i$, as an element of $\mathbb{C}P^2$, could be
rewritten as a scalar multiple $p_i \sim \frac{1}{1 + {\mathrm i}
\beta}\;p_i = \re\,p_i$, showing that $p_i$ is totally real.  In
other words, $p_i$ would lie in $\mathbb{R}P^2 \subset
\mathbb{C}P^2$. In particular, $p_i$ would be its own complex
conjugate, and we can hardly expect a single point $p_i =
\overline{p_i}$ to determine a unique line. Fortunately this case
cannot occur, because $Q(x,y,z) = x^2 + y^2 + z^2 = 0$ admits no
real solutions.

So let $L_i$ denote the unique line $a_i x + b_i y + c_i z = 0$
containing the conjugate pair $\{p_i, \overline{p_i}\}$. More
precisely, let $L_i = a_i x + b_i y + c_i z$ be the unique (modulo
rescaling) real linear polynomial whose roots include both $p_i$
and $\overline{p_i}$.  The desired decomposition
(\ref{MainTheoremEquation}) becomes
\begin{eqnarray}
\label{MainTheoremEquationBis}
  P \; = \; \lambda\;L_1 L_2 \cdots L_\ell \; + \; Q\cdot R.
\end{eqnarray}
To prove that this equality holds, we again turn to B\'ezout's
Theorem.  First recall that the complex curve $P = 0$ intersects
the complex curve $Q = 0$ in precisely the $2\ell$ points $\{p_1,
\overline{p_1}, \dots, p_\ell, \overline{p_\ell}\}$.  By
construction, the product curve $L_1 L_2 \cdots L_\ell = 0$ also
intersects $Q = 0$ in those same $2\ell$ points, and by B\'ezout's
Theorem there are no other points of intersection.  Now pick any
other point $q \in \{Q = 0\}$ and define $\lambda$ to be the ratio
\begin{equation}
\label{DefinitionLambda}
  \lambda = \frac{P(q)}{L_1(q) L_2(q) \cdots L_\ell(q)}.
\end{equation}
Write a new polynomial
\begin{equation}
\label{DefinitionOfF}
  F \equiv P - \lambda L_1 L_2 \cdots L_\ell.
\end{equation}
This new polynomial $F$ has degree $\ell$, yet has zeros at the
$2\ell + 1$ distinct points $\{\;q,\; p_1, \overline{p_1}, \dots,
p_\ell, \overline{p_\ell}\} \subset Q$.  In other words, the
complex curve $F = 0$ intersects the complex curve $Q = 0$ at (at
least) $2\ell + 1$ distinct points.  By B\'ezout's Theorem the
polynomials $F$ and $Q$ must share a common factor;  because $Q$
is irreducible the common factor must perforce be $Q$ itself. Thus
we may factor $F$ as
\begin{equation}
\label{Factorization}
  F = Q \cdot R
\end{equation}
for some remainder term $R$.  Combining (\ref{DefinitionOfF}) and
(\ref{Factorization}) yields the desired decomposition
(\ref{MainTheoremEquationBis}).

Let us now prove that $\lambda$ is real.  In light of the
factorization (\ref{Factorization}), the polynomial $F$ is clearly
zero on the whole complex curve $Q = 0$. In particular, for the
point $q$ chosen earlier,
\begin{equation}
\label{Step1}
  F(q) = F(\bar{q}) = 0.
\end{equation}
On the one hand
\begin{equation}
\label{Step2}
  F(\bar{q}) = P(\bar{q}) - \lambda L_1(\bar{q}) L_2(\bar{q}) \cdots L_\ell(\bar{q}).
\end{equation}
On the other hand, because $P$ and the $L_i$ all have real
coefficients,
\begin{equation}
\label{Step3}
  \overline{F(q)} = P(\bar q) - \overline{\lambda}
     L_1(\bar q) L_2(\bar q) \cdots L_\ell(\bar q)
\end{equation}
Comparing (\ref{Step1}), (\ref{Step2}) and (\ref{Step3}), and
recalling that $q$ was chosen to ensure $L_i(q) \neq 0$, proves
that $\lambda = \bar\lambda$, in other words, $\lambda$ is real.

An elementary argument then shows that for all real values of $x$,
$y$ and $z$, $R(x,y,z)$ = $\overline{R(x,y,z)}$ =
$\overline{R}(x,y,z)$, implying that the coefficients of the
polynomial $R$ must all be real.

This completes the proof of the existence part of Theorem~1.

\vskip0.5cm

Let us now prove that the decomposition (
\ref{MainTheoremEquation}) is unique.  Assume we have two
decompositions
\begin{eqnarray}
\label{TwoDecompositions}
  P(x,y,z) &=& \lambda\phantom{'}  \; L_1 L_2 \cdots L_\ell
               \;+\; Q\cdot R \nonumber\\
           &=& \lambda' \; L'_1 L'_2 \cdots L'_\ell
               \;+\; Q\cdot R'.
\end{eqnarray}
Our goal is to show that each linear factor $L'_{i'}$ in the
second decomposition occurs as a factor $L_i$ in the first
decomposition as well, modulo a possible rescaling.  A given line
$L'_{i'} = 0$ intersects the quadratic $Q = 0$ in a pair of
conjugate points $p$ and $\bar p$.  Because $p$ and $\bar p$
satisfy both $L'_{i'} = 0$ and $Q = 0$, they satisfy $P = 0$ as
well.  Turning our attention to the first decomposition, because
$p$ and $\bar p$ satisfy both $P = 0$ and $Q = 0$, they satisfy
$L_1 L_2 \cdots L_\ell = 0$ as well.  Hence $p$ must satisfy one
of the lines $L_i = 0$, and because the line's coefficients are
real, $\bar p$ must satisfy that same line.  But we saw earlier
that a pair of conjugate points $p$ and $\bar p$ determines a
unique line modulo normalization (recall the essentially unique
solution to Equations~(\ref{LineEquations})).  Therefore $L_i$ is
a constant multiple of $L'_{i'}$, and if the coefficients of each
have been normalized to length one, then $L_i = \pm L'_{i'}$. This
proves the uniqueness of the factorization.

If we evaluate the two decompositions~(\ref{TwoDecompositions}) on
the complex curve $Q = 0$ we get
\begin{equation}
    \lambda  \; L_1 L_2 \cdots L_\ell
  = \lambda' \; L'_1 L'_2 \cdots L'_\ell
\end{equation}
proving that if the coefficients of the $L_i$ and the $L'_{i'}$
are consistently normalized, then $\lambda = \lambda'$. It then
follows easily that $R = R'$ as well.

This completes the proof that the decomposition (
\ref{MainTheoremEquation}) is unique, thus completing the proof of
Theorem 1.

\section{Computational Considerations}

The proof presented in Section~\ref{SectionProof} is almost
constructive, but not quite.  It relies on B\'ezout's Theorem for
the existence of the root pairs $\{p_1, \overline{p_1}, \dots,
p_\ell, \overline{p_\ell}\}$ but does not say how to find them.
This section fills the gap.

The key observation is that the quadratic curve $Q = x^2 + y^2 +
z^2 = 0$ is topologically a 2-sphere.  More to the point, it is a
copy of the complex projective line $\mathbb{C}P^1$, which happens
to be homeomorphic to the 2-sphere.  Let us parameterize the curve
$Q = 0$ as
\begin{equation}
\label{Parameterization}
  (\,i\,(u^2 - v^2), \; -2i \, u v, \; u^2 + v^2)
\end{equation}
where $u$ and $v$ are homogeneous coordinates in $\mathbb{C}P^1$.
Clearly the mapping (\ref{Parameterization}) takes all points
$(u,v) \in \mathbb{C}P^1$ to the curve $Q = 0$, by construction.
The question is, which of those points happen to satisfy the given
polynomial $P$ as well?  Write
\begin{equation}
\label{PonCP1}
  P(x,y,z) = P(\,i\,(u^2 - v^2), \; -2i \, u v, \; u^2 + v^2)
\end{equation}
to express $P$ as a function on $\mathbb{C}P^1$.

If $v \neq 0$, then $(u,v)$ and $(\frac{u}{v}, 1)$ represent the
same point in $\mathbb{C}P^1$. If we define $\alpha \equiv
\frac{u}{v}$ then expression (\ref{PonCP1}) effectively reduces to
a polynomial in a single variable,
\begin{equation}
\label{PonCP1bis}
  P(x,y,z) = P(\,i\,(\alpha^2 - 1), \; -2i \, \alpha, \; \alpha^2 + 1)
\end{equation}
The roots of this polynomial are the desired root pairs $\{p_1,
\overline{p_1}, \dots, p_\ell, \overline{p_\ell}\}$.

If, on the other hand, $v = 0$, then $(u,v) = (u,0) \sim (1,0)$.
Thus $(u,v) = (1,0)$ may represent an additional root, which would
not be found as a root of $P(\alpha)$ in (\ref{PonCP1bis}).

Once we have found the parameters $(u,v)$ for all $2\ell$ roots of
$P$, the easiest way to group them into conjugate pairs is to
observe that the parameterization (\ref{Parameterization}) maps
``antipodal points'' $(u,v), (-\bar v, \bar u) \in \mathbb{C}P^1$
to conjugate points $(x, y, z), (\bar x, \bar y, \bar z) \in
\mathbb{C}P^2$. In other words, $(\alpha,1)$ and $(-1,\bar\alpha)
\sim (-1/\bar\alpha,1)$ map to a pair of conjugate points in
$\mathbb{C}P^2$.

\section{Examples}

To illustrate how the algorithm works in practice, let us apply
Theorem~1 to several concrete examples.

\subsection{Toy Quadrupole}
\label{SubsectionToyQuadrupole}

Consider the quadratic polynomial
\begin{equation}
\label{Example1}
  P(x,y,z) = xy + yz + zx - x^2 - z^2.
\end{equation}

First dismiss the special case $(u,v) = (1,0)$ by noting that the
parameterization~(\ref{Parameterization}) maps $(u,v) = (1,0)$ to
$(x,y,z)= (i,0,1)$ for which (\ref{Example1}) gives $P(i,0,1) = i
\neq 0$.

Now consider the general case, for which
Equation~(\ref{PonCP1bis}) becomes
\begin{equation}
  i \alpha^4 + 2(1-i)\alpha^3 - 4\alpha^2 - 2(1+i)\alpha - i = 0
\end{equation}
with roots
\begin{eqnarray}
 \alpha_1 &=& 1 + \sqrt{2} \nonumber\\
 \alpha_2 &=& 1 - \sqrt{2} \nonumber\\
 \alpha_3 &=& i(1 + \sqrt{2}) \nonumber\\
 \alpha_4 &=& i(1 - \sqrt{2}).
\end{eqnarray}
corresponding, respectively, to the four points of
$\mathbb{C}P^2$,
\begin{eqnarray}
           p_1  &=& (1, -1, -i\sqrt{2}) \nonumber\\
 \overline{p_1} &=& (1, -1, +i\sqrt{2}) \nonumber\\
           p_2  &=& (-i\sqrt{2}, 1, -1) \nonumber\\
 \overline{p_2} &=& (+i\sqrt{2}, 1, -1).
\end{eqnarray}
Solving the line equations~(\ref{LineEquations}) converts the
preceding two pairs of conjugate points to the two lines
\begin{eqnarray}
  L_1 &=& \sqrt{\half}\,x + \sqrt{\half}\,y \phantom{+ \sqrt{\half}\,z} = 0  \nonumber\\
  L_2 &=& \phantom{\sqrt{\half}\,x +} \sqrt{\half}\,y + \sqrt{\half}\,z = 0,
\end{eqnarray}
which give us the two multipole vectors $(\sqrt\half, \sqrt\half,
0)$ and $(0, \sqrt\half, \sqrt\half)$.

To find the correct $\lambda$, evaluate
Equation~(\ref{DefinitionLambda}) for, say, $q = (1,i,0)$, giving
\begin{eqnarray}
 \lambda = \frac{P(q)}{L_1(q) L_2(q)}
 = \frac{-1 + i}{-\half + \frac{i}{2}}
 = 2.
\end{eqnarray}
Of course any other choice for $q$ would have given the same
answer $\lambda = 2$, just so we make sure $q$ lies on the curve
$Q(q) = x^2 + y^2 + z^2 = 0$ and exclude $q \in \{p_1,
\overline{p_1}, p_2, \overline{p_2}\}$.

We may now write down the polynomial $F$ from
Equation~\ref{DefinitionOfF}, namely
\begin{eqnarray}
  F &=& P - \lambda L_1 L_2 \nonumber\\
    &=& (xy + yz + zx - x^2 - z^2) \nonumber\\
    & &   - 2(\sqrt{\half}x + \sqrt{\half}y)
          (\sqrt{\half}y + \sqrt{\half}z) \nonumber\\
    &=& -x^2 - y^2 - z^2,
\end{eqnarray}
and divide by $Q = x^2 + y^2 + z^2$ to get the remainder term $R =
F/Q = -1$.  Thus the final decomposition promised by Theorem 1
becomes
\begin{eqnarray}
  & & xy + yz + zx - x^2 - z^2 \nonumber\\
  &=& 2(\sqrt{\half}x + \sqrt{\half}y)
          (\sqrt{\half}y + \sqrt{\half}z) \nonumber\\
  & &     \quad + \quad (x^2 + y^2 + z^2)(-1).
\end{eqnarray}

\subsection{Toy Octopole}
\label{SubsectionToyOctopole}

The cubic polynomial
\begin{equation}
\label{Example2}
  P(x,y,z) = x^2 y + y^3
\end{equation}
illustrates some non-generic behavior which may arise, namely the
possibilities of {\it (a)} a ``missing root'' and {\it (b)}
multiple roots. We will follow the same algorithm as in
Section~\ref{SubsectionToyQuadrupole}, pointing out only the
differences.

The first difference is that the special case $(u,v) = (1,0)$,
corresponding to $(x,y,z)= (i,0,1)$, is indeed a root of $P$ in
(\ref{Example2}).  So we record that root and proceed onward in
search of the other roots.

The next difference we encounter is that the polynomial
\begin{equation}
\label{Alpha2}
  2i \alpha^5 + 4i \alpha^3 + 2i \alpha = 0.
\end{equation}
has degree only 5, not degree $2\ell = 2\cdot 3 = 6$ as one
expects in the generic case.  Happily, this polynomial's five
roots supplement the one exceptional root $(i,0,1)$ we found in
the previous paragraph, to give the required total of six roots.
In other words, the unexpectedly low degree of the polynomial is
intimately tied to the existence of the exceptional root
$(i,0,1)$.

The roots of (\ref{Alpha2}) turn out to be $\{-i, i, -i, i, 0\}$.
Unlike more generic polynomials, this one has multiple roots,
implying a repeated factor in the product $L_1 L_2 L_3$.
Specifically, those five roots correspond to
\begin{eqnarray}
           p_1  &=& (+i, 1, 0) \nonumber\\
 \overline{p_1} &=& (-i, 1, 0) \nonumber\\
           p_2  &=& (+i, 1, 0) \nonumber\\
 \overline{p_2} &=& (-i, 1, 0) \nonumber\\
           p_3  &=& (-i, 0, 1),
\end{eqnarray}
and then the one exceptional root $(i,0,1)$ completes the pattern
\begin{eqnarray}
 \overline{p_3} &=& (+i, 0, 1).
\end{eqnarray}

From here the algorithm is routine.  The lines are
\begin{eqnarray}
  L_1 = z = 0\phantom{,}  \nonumber\\
  L_2 = z = 0\phantom{,}  \nonumber\\
  L_3 = y = 0,
\end{eqnarray}
the scalar multiple is $\lambda = -1$, and the final factorization
is
\begin{eqnarray}
  x^2 y + y^3 = -1(y)(z)(z) + (x^2 + y^2 + z^2)(y).
\end{eqnarray}

\subsection{WMAP Quadrupole and Octopole}

Our first task here is to convert a given set of coefficients
$a_{\ell m}$ to a homogeneous harmonic polynomial. Converting the
standard spherical harmonics $Y_{\ell}^m$ to polynomials in $x$,
$y$ and $z$ is easy. For example, for the quadrupole,
\begin{eqnarray}
  Y_2^{-2} &=& \sqrt{\frac{15}{32\pi}} \; \sin^2 \theta \; e^{-2i\varphi}
      = \sqrt{\frac{15}{32\pi}} \; (x - i y)^2 \nonumber\\
  Y_2^{-1} &=& \sqrt{\frac{15}{8\pi}} \; \sin\theta \cos\theta \; e^{-i\varphi}
      = \sqrt{\frac{15}{8\pi}} \; (x - i y)z \nonumber\\
  Y_2^{\phantom{-}0} &=& \sqrt{\frac{5}{16\pi}} \; (3 \cos^2\theta - 1)
      = \sqrt{\frac{5}{16\pi}} \; (3 z^2 - 1) \nonumber\\
      &=& \sqrt{\frac{5}{16\pi}} \; (3 z^2 - (x^2 + y^2 + z^2)) \nonumber\\
      &=& \sqrt{\frac{5}{16\pi}} \; (-x^2 - y^2 + 2 z^2) \nonumber\\
  Y_2^{\phantom{-}1} &=& -\sqrt{\frac{15}{8\pi}} \; \sin\theta \cos\theta \; e^{i\varphi}
      = -\sqrt{\frac{15}{8\pi}} \; (x + i y)z \nonumber\\
  Y_2^{\phantom{-}2} &=& \sqrt{\frac{15}{32\pi}} \; \sin^2 \theta \; e^{2i\varphi}
      = \sqrt{\frac{15}{32\pi}} \; (x + i y)^2
\end{eqnarray}
Using the coefficients $a_{2,m}$ for the DQ-corrected Tegmark map
of the first-year WMAP quadrupole gives
\begin{eqnarray}
  P(x,y,z) &=& -0.01262739 \; x^2 + 0.02302019 \; x y \nonumber\\
           &+&  0.00677625 \; y^2 + 0.00950698 \; x z \nonumber\\
           &+&  0.01064014 \; y z + 0.00585113 \; z^2.
\end{eqnarray}
Following the same algorithm, as illustrated in Sections
\ref{SubsectionToyQuadrupole} and \ref{SubsectionToyOctopole}, we
get the polynomial
\begin{eqnarray}
  (0.01847852 - i\;0.00950698) \; \phantom{\alpha^0}\nonumber\\
  - (0.04604038 + i\;0.02128027) \; \alpha^{\phantom{1}}\nonumber\\
  - 0.04065752\phantom{)} \; \alpha^2\nonumber\\
  + (0.04604038 - i\;0.02128027) \; \alpha^3 \nonumber\\
  + (0.01847852 + i\;0.00950698) \; \alpha^4
\end{eqnarray}
leading to the lines
\begin{eqnarray}
  L_1 = \phantom{-}0.939660\;x + 0.187066\;y + 0.286437\;z = 0\phantom{.}  \nonumber\\
  L_2 = -0.437088\;x + 0.792820\;y + 0.424724\;z = 0.
\end{eqnarray}
Converting the coefficients of these lines to spherical
coordinates gives multipole vectors
\begin{eqnarray}
  \hat{v}^{(2,1)} &=& (\phantom{0}11.26^\circ, 16.64^\circ)  \nonumber\\
  \hat{v}^{(2,2)} &=& (118.87^\circ, 25.13^\circ)
\end{eqnarray}
in full agreement with those that CHS found using their tensor
algorithm (Equation~3 of \cite{Schwarz}).

An analogous computation for the octopole yields multipole vectors
$\hat{v}^{(3,1)}$, $\hat{v}^{(3,2)}$ and $\hat{v}^{(3,3)}$, again
in full agreement with those reported in Equation~3 of
\cite{Schwarz}.

\section{How well do the WMAP quadrupole and octopole align?}

Following \cite{Schwarz}, we define the {\it quadrupole plane}
normal vector
\begin{eqnarray}
  w^{2,1,2} \equiv \hat{v}^{(2,1)} \times \hat{v}^{(2,2)}
\end{eqnarray}
and the three {\it octopole plane} normal vectors
\begin{eqnarray}
  w^{3,1,2} \equiv \hat{v}^{(3,1)} \times \hat{v}^{(3,2)}\phantom{.}\nonumber\\
  w^{3,2,3} \equiv \hat{v}^{(3,2)} \times \hat{v}^{(3,3)}\phantom{.}\nonumber\\
  w^{3,3,1} \equiv \hat{v}^{(3,3)} \times \hat{v}^{(3,1)}.
\end{eqnarray}
Still following \cite{Schwarz}, we judge the alignment of the
quadrupole plane with the three octopole planes via the dot
products
\begin{eqnarray}
  A_1 = | w^{2,1,2} \cdot w^{3,2,3} | \phantom{.}\nonumber\\
  A_2 = | w^{2,1,2} \cdot w^{3,3,1} | \phantom{.}\nonumber\\
  A_3 = | w^{2,1,2} \cdot w^{3,1,2} | .
\end{eqnarray}
Finally, in contrast to \cite{Schwarz} (which in its preprint form
contained some statistical flaws), we let the sum
\begin{eqnarray}
  S = A_1 + A_2 + A_3
\end{eqnarray}
provide a numerical measure of how well the quadrupole plane
aligns with the octopole planes.  For the DQ-corrected Tegmark
map, the sum evaluates to $S_0 = 2.395$.

To judge how unusually large $S_0$ is, we evaluated $S$ for 100000
random quadrupoles and octopoles, and found that only 118 trials
produced $S > S_0$.  This 99.9\% confidence level, while weaker
than the incorrectly stated results of \cite{Schwarz}, is
completely consistent with Huterer and Starkman's revised
statistical analysis \cite{Glenn}.

Like Schwarz, Starkman, Huterer and Copi, we find this result
astonishing.  In particular we find it difficult to believe that
the quadrupole and octopole align so well merely by chance.
Whether the alignment is imposed by the global topology of a small
finite universe, or whether it is due to some previously unknown
solar system effect, or whether it is merely the result of an
error in the collection and/or processing of the WMAP data,
remains to be seen.

In the meantime, we emphasize that our simulations use an entirely
different algorithm from that of \cite{Copi,Schwarz} as well as
completely independent computer code.  This effectively rules out
the possibility of error in computing the 1-in-1000 estimate,
forcing us to take that estimate quite seriously.

We should point out that our Monte Carlo simulations chose random
quadrupoles and octopoles independently of each other, relative to
spherically symmetric distributions on the spaces of all spherical
harmonics of degree 2 and 3, respectively.  In other words, we
used independent Gaussian coefficients $a_{\ell m}$.

\section{An Open Question}

Corollary 2 motivates a broader question, first raised by Copi,
Huterer and Starkman \cite{Glenn2}.  One would like to decompose
an arbitrary real-valued function $f: S^2 \rightarrow R$, for
example the temperature function on the microwave sky, as a sum $f
= \sum_{\ell = 0}^\infty \left(\lambda_\ell \prod_{i=1}^\ell
L_{\ell,i}\right)$. In other words, this approach would bypass the
spherical harmonics entirely, and instead write the function $f$
directly as the sum of totally factored polynomials $\lambda_\ell
\prod_{i=1}^\ell L_{\ell,i}$, one for each degree $\ell$.

Corollary 2 almost makes such a factorization possible.  For
example, if we approximate the microwave sky temperature by the
sum of its first 837 multipoles, $T = \sum_{\ell = 0}^{837}
\sum_{m = -\ell}^\ell a_{\ell,m} Y_\ell^m$, then Corollary~2 lets
us re-express it as $T = \sum_{\ell = 0}^{837} \left(\lambda_\ell
\prod_{i=1}^\ell L_{\ell,i}\right)$.  The question is, what
happens when we add in the $838^{th}$ spherical harmonic?  For
sure we will add an $838^{th}$ term $\lambda_{838}
\prod_{i=1}^{838} L_{838,i}$ to our factored-polynomial
decomposition.  And almost surely the $836^{th}$ term will change
significantly as it inherits the remainder $R$ from the newly
added $838^{th}$ term.  But what about the lower order terms? Will
the second, fourth and sixth terms remain stable?  Or will they
swing wildly every time we add a new high-order term to the
series?  In other words, is the decomposition stable?

\end{document}